\documentstyle[pra,graphicx,aps]{revtex}

\begin{document}


\title{
	Threshold and non-linear behavior of lasers of $\Lambda$ and V - 
	configurations
}
\author{
		Gennady A. Koganov\thanks{quant@bgumail.bgu.ac.il} and Reuben Shuker
}
\address{Physics Department, Ben Gurion University of the Negev, 
P.O.Box 653, Beer Sheva 84105, Israel}
\maketitle 

\begin{abstract}
Dynamic properties of closed three level laser systems are investigated. Two schemes of pumping  - $\Lambda$ and V - are considered. It is shown that the non-linear behavior of the photon number as a function of pump both near and far above threshold 
is crucially different for these two configurations. In particular, it is found that in the high pump regime laser can turn off in a phase-transition-like manner in both $\Lambda$ and V schemes. \\ 
\end{abstract}


\paragraph*{}
The interest in the dynamic behavior of lasers and its analogy to phase 
transition phenomenon have been renewed over recent years mainly due to 
experimental realization of microlasers [\ref{Feld}]. The reason for this is a 
possibility of lowering of the threshold pump needed to start the lasing process. 
The value of the threshold pump is mostly determined by the fraction $\beta$ of 
spontaneously emitted photons directed into the lasing mode [\ref{beta}]. 
It has been shown that in the limit when all spontaneous photons are emitted 
into the lasing mode (which corresponds to the spontaneous emission factor 
$\beta = 1$), the laser becomes a "thresholdless" device [\ref{RC}]. Such lasers 
are also referred to as cavity-QED lasers. 

\paragraph*{}
Mu and Savage [\ref{Savage}] pointed out that when the pumping excites the lower 
lasing level the number of photons can decrease with pump, and the laser can even 
turn off at strong enough pump rate. 
Recently we have shown that the type of nonlinear dependence of the photon number 
upon the pump rate is very sensitive to the type of pumping used in a laser 
[\ref{we1}]. Figure 1 illustrates the difference between the 
two types of pumping on the example of three level atom. 
In the scheme shown 
in fig. 1a an atom upon emitting a photon decays from the lower lasing 
state $|0>$ 
to the ground state $|2>$ with the decay rate $\gamma_{02}$, afterwards the pumping  
excites the 
atom to the upper lasing state $|1>$ with the rate $\gamma_{21}$ . Thus in this case 
the rate 
$\gamma_{21}$ plays a role of the pump rate. In the 
scheme shown in fig. 1b on the other hand, the pumping first excites the atom, with 
the 
rate $\gamma_{02}$ from the lower lasing state $|0>$ which is the ground state in 
this case, to the excited state $|2>$ which is then deplited with the decay 
rate $\gamma_{21}$ to the upper lasing state $|1>$. Now the role of the pump 
rate is 
played by the rate 
$\gamma_{02}$. We  refer to the schemes depicted in figs. 
1a and 1b as $\Lambda$ and V configurations, respectively. One has to distinct, however, 
between these $\Lambda$ and V schemes and those discussed in the literature on lasing 
without inversion [\ref{LWI}]. The main feature of our V-type is that the pumping is 
used to excite the 
lower lasing state $|0>$ directly. It is this property of the V-type schemes that 
makes their dynamic 
behaviour crucially different from that of the $\Lambda$ -type schemes.
\paragraph*{}€ 
In the present paper we study the dynamic behavior of the lasers with $\Lambda$ 
and V  types of pumping using modified Maxwell-Bloch equations wich include a 
term describing spontaneous emission into the lasing mode. We follow gradual 
transition 
from the threshold-like to thresholdless behaviour of $\Lambda$- type scheme. 
We also show that if the saturation parameter (or cooperativity parameter) is larger 
than some critical value the lasing is not possible altogether. This implies a 
restriction to the 
spontaneous emission factor $\beta$ which has a minimal value. The V-type 
scheme always has a threshold, however the  pump parameter is restricted by 
its maximal value which is a critical point as at this point the lasing is 
broken down in a phase-transition-like manner. For such a scheme the factor 
$\beta$ is not a parameter because of its dependence on the pump rate. 
\paragraph*{}€
We start from the set of modified Maxwell-Bloch equations which describe both 
$\Lambda$ and V-type schemes shown in fig. 1. These equations are 
derived by writing down an exact  master equation for the atoms+field density 
matrix in divergent form and omitting the fluctuation terms.

\begin{eqnarray}
	\dot{n}=-2\kappa n+Ngx & € & €
	\label{eq1}  \\
	\dot{x}=-\gamma_{\perp} x+2g[(n+1)\rho_{11}-n\rho_{00}]€ & €
	\label{eq2}  \\
	\dot{\rho}_{11}=-\gamma_{10}\rho_{11}+\gamma_{21}\rho_{22}-gx
	\label{eq3}  \\
	\dot{\rho}_{00}=-\gamma_{02}\rho_{00}+\gamma_{10}\rho_{11}+gx
	\label{eq4}  \\
	\rho_{00}+\rho_{11}+\rho_{22}=1€ & € & €
	\label{eq5}
\end{eqnarray}€

with $x \equiv z^{*}\rho_{10} + c.c.$, N is the total number of atoms, 
g is the coupling constant, $\kappa$ is the cavity decay rate, $n=z^{*}z$ is the 
number of 
photons in the cavity. 
Equations (\ref{eq1})-(\ref{eq5}) differ from the standard Maxwell-Bloch equations 
by the 
factor n+1 instead of n in eq. (\ref{eq2}), which takes account for spontaneous 
emission into the lasing mode. Such a factor occasionaly appears in rate 
equations [\ref{Sigman}], however eqs. (\ref{eq1})-(\ref{eq5})  provide a more complete description 
as their 
validity is not restricted by the approximation $\gamma_{\perp} \gg \gamma_{ij}$ 
used to derive the rate equations.
It follows from the derivation of the relaxation rates [\ref{gammas}] that the 
transversal relaxation rate is related to the other rates by the following 
important formula
\begin{equation}
\gamma_{\perp}=\frac{1}{2}(\gamma_{10}+\gamma_{02}+\gamma_{col}€)€€€
	\label{eq6}
\end{equation}€

with $\gamma_{col}$ being the rate of collisional dephasing. 
This brings about additional dependence of the photon number upon the rate 
$\gamma_{02}$ which in the case of V- configuration plays a role of a pump rate.
The steady state solution for the  number of photons n  is

\begin{equation}
	n=\frac{1}{2}(b+\sqrt{b^{2}+4c})
	\label{eq7}
\end{equation}€

with

\begin{eqnarray}
	b=\frac{\lambda 	
	(\alpha_{2}-1)-S(\alpha_{2}+1+\eta)(\alpha_{2}\alpha_{1}+\alpha_{2}+\alpha_{1})
	-\alpha_{1}-\alpha_{2}}{\alpha_{2}+2\alpha_{1}}  ,\:\;
	c=\frac{\lambda\alpha_{1}\alpha_{2}}{\alpha_{2}+2\alpha_{1}}
	\label{eq8}  \\
	\lambda\equiv \frac{N\gamma_{10}}{2\kappa}, 
	\;\;\;S\equiv\frac{\gamma_{10}^{2}}{4g^{2}},\;\;\; \alpha_{1}\equiv 
	\frac{\gamma_{21}}{\gamma_{10}},\;\;\;\alpha_{2}\equiv 
	\frac{\gamma_{02}}{\gamma_{10}}, \;\;\;\eta\equiv 
	\frac{\gamma_{col}}{\gamma_{10}}
	\label{eq8a}
\end{eqnarray}€

Consider first the $\Lambda$-configuration (fig. 1a). In this case the 
role of the pump parameter is played by $\alpha_{1}$, so 
$P_{1}=\alpha_{1}$. In fig. 2 the number of photons is plotted as a fuction of 
the pump parameter $P_{1}$ for various values of the saturatoin 
parameter S. 
When S is not too large and the pump parameter is small 
enough ($P_{1}<1$) one can observe (i) threshold kinks which tend 
to disappear as S is approaching zero and (ii) linear dependence of the 
photon number on the pump above threshold.  In this regime our results are similar to those 
obtained by other authors [\ref{RC}, \ref{Jin}, \ref{Yamamoto}]. However the 
picture changes in 
two ways when either $P_{1}$ or S become large enough. The
photon number saturates at large values of the pump which is not surprizing 
since when the pumping excites the groung state $|2>$ very fast there is a 
bottleneck at the transition $|0>-|2>$ which prevents further growth of the 
photon number. This kind of non-linearity (saturation) was discussed by Hart and Kennedy  
[\ref{Hart,Kenedy}] under similar conditions. Note that the curves for different values of the saturation 
parameter S have different saturation limits at $P_{1}\rightarrow\infty$. Another 
change occurs when the saturation parameter S approaches some critical value 
$S_{cr}$ ($7.44*10^{5}$ for the set of parameters used in fig. 2). The 
kinks become less defined with increasing  S and at the same time the 
saturation  photon number decreases drastically.  At $S = S_{cr}$ (curve 
xii in fig. 2) 
the kinks disappear and further insreasing of S results in a no lasing regime.
\paragraph*{}
Consider a 
threshold condition for the $\Lambda$- scheme which can be derived from eqs. 
(\ref{eq1})-(\ref{eq5}) 
with n+1 replaced by n in eq. (\ref{eq2}), i.e. from the standard Maxwell-
Bloch equations, as a condition for the existence of a positive steady state 
solution for the photon number. Then the threshold condition reads

\begin{equation}
	P_{1} > P_{1thr}=\frac{\alpha_{2}S(\alpha_{2}+1+\eta)}{\lambda(\alpha_{2}-1)-
	S(\alpha_{2}+1)(\alpha_{2}+1+\eta)}
	\label{eq9}
\end{equation}

One can make two major observations from eq. (\ref{eq9}). The first one is well 
known fact that the 
threshold pump decreases with S and in the limit $S\rightarrow0$, when the kinks in 
the curves desappear, $P_{1thr}\rightarrow 0$, i.e. the laser becomes
thresholdless [\ref{RC}, \ref{Jin}, \ref{Yamamoto}]. Secondly, the saturation 
parameter S is restricted by its maximal value

\begin{equation}
	S < S_{max}=\frac{\lambda 
	(\alpha_{2}-1)}{(\alpha_{2}+1)(\alpha_{2}+1+\eta)}€
	\label{eq10}
\end{equation}

which for the set of parameters used in fig. 2 corresponds exactly to the 
curve xii. Therefore the curves xii-xv describe no-lasing regime. In this sense
the value $S_{max}$ is a critical one as it separates the regimes with and 
without lasing. The physical meaning of unequality (\ref{eq10}) 
becomes apparent if we rewrite it in the following equivalent 
form (provided $\gamma_{02}>>\gamma_{10}$) 
$\gamma_{02}+\gamma_{col}<2Ng^{2}/\kappa$ which, together with eq. 
(\ref{eq6}), implies a restriction on the time $T_{2}$. 
\paragraph*{}
Another interpretation can be obtained if we relate the parameters defined in 
eq. (\ref{eq8a}) with the spontaneous emission factor $\beta$ which results in the 
following 

\begin{equation}
	\beta=\frac{1}{1+S(1+\alpha_{2}+\eta)},\;\;\;\;\;\;\;
	\beta_{min}=\frac{1}{1+\lambda\frac{\alpha_{2}-1}{\alpha_{2}+1}}
	\label{eq11}
\end{equation}

Thus the limit $S\rightarrow 0$ corresponds to an ideal QED-laser 
($\beta=1$) in 
which case all spontaneous photons are directed into the lasing mode. Note 
that there exists a minimal value of $\beta$ which means that the laser can 
not operate if the portion of spontaneously emitted photons directed into the 
lasing mode is less than $\beta_{min}$. If 
$\alpha_{2}$ and $\lambda >>1$,  $\beta_{min} = 1/\lambda$. The physical meaning 
of the critical point 
$\beta=\beta_{min}$ is illustrated in fig. 3 where the
saturation photon number $n_{sat}$ (which is calculated as a limit of the photon number n at $P_{1}\rightarrow\infty$) is plotted as a function of $\beta$ (S) parameter.

\paragraph*{}
Now consider the V-scheme shown in fig. 1b. In this case the pumping 
excites the lower lasing state $|0>$ which is also the ground state 
and so the rate $\gamma_{02}$ plays a role of the pump rate.  
Therefore it is reasonable to introduce the pump parameter 
$P_{2}=\alpha_{2}$. The dependence of the photon number n upon this 
pump parameter is essentially different than that of the 
$\Lambda$-scheme due to relation between the transversal relaxation 
rate $\gamma_{\perp}$ and the pump parameter $P_{2}$ following from eq. 
(\ref{eq6}) and the definition of the pamp parameter $P_{2}$, namely

\begin{equation}
	\gamma_{\perp}=\frac{\gamma_{10}}{2}(1+P_{2}+\eta)
	\label{eq12}
\end{equation}

which brings about additional dependence on the pump parameter. This 
differnce is clearly seen in fig. 4 where the photon number n as well 
as the inversion $D=\rho_{11}-\rho_{00}$ are plotted as functions of 
the pump parameter $P_{2}$ for various values of the saturation 
parameter S. Now as the pump parameter increases the photon number 
grows to some critical value, at which it begins to decrease. The 
curves $n(P_{2})$ have two kinks; the first kink corresponds to the 
threshold point whereas the second one - to the break point at which 
lasing ceases. Note that in contrast to the $\Lambda$-scheme (i) the 
values of the threshold pump for all curves in fig. 4a are closed to 1 
and (ii) the kinks do not disappear at $S\rightarrow 0$ ($\beta \rightarrow 
1$), i.e. the threshold exists even in an ideal QED-laser. 

\paragraph*{}
The two critical points can be obtained from semiclassical steady 
state equations by solving the inequality $n>0$, as it has been done 
for the $\Lambda$-scheme, which results in the following

\begin{equation}
	P_{2thr}<P_{2}<P_{2max}
	\label{eq13}
\end{equation}

with

\begin{eqnarray}
	 & € & 
	 €P_{2thr}=1+\frac{S}{\lambda}\frac{(1+2\alpha_{1})(2+\eta)}{\alpha_{1}}
	\label{eq14}  \\
	€ & € & €P_{2max}=\frac{\lambda}{S}\frac{\alpha_{1}}{1+\alpha_{1}}
	\label{eq15}
\end{eqnarray}

Approximate equations (\ref{eq14}) and (\ref{eq15}), obtained by 
expansion of exact expressions for $P_{2thr}$ and $P_{2max}$ at 
$S<<\lambda$, work pretty well for any reasonable values of the 
involved parameters. The presence of the second critical point 
$P_{2max}$ is caused by the dependence of the transversal relaxation 
rate $\gamma_{\perp}$ on the pump parameter $P_{2}$ (see eq. 
(\ref{eq12})). In fact, eq. (\ref{eq15}) gives rise to the same 
restriction for the depletion rate $\gamma_{02}$ as in the case of the 
$\Lambda$-scheme, this time, however, this is the restriction on the 
pump rate. This causes the reason for breaking down of lasing at 
the point $P_{2}=P_{2max}$. 

\paragraph*{}
Another interesting feature of the V-scheme is that the spontaneous 
emission factor $\beta$ is not a parameter as it depends upon the pump 
parameter $P_{2}=\alpha_{2}$ (see eq. (\ref{eq11})). Therefore increasing of the pump 
results in decreasing of $\beta$ which is restricted by its minimal 
value. In this sense the breaking point $P_{2}=P_{2max}$ in the 
V-scheme and the point $S=S_{max}$ in the $\Lambda$ -scheme have the 
same physical origin - both stem from the restriction on $T_{2}$. Figure 5 summarizes the results 
for the V-scheme by plotting the photon number, the polarization and the inversion as a function of both the pump parameter $P_{2}$ and the spontaneous emission factor $\beta$ in one plot. One sees what happens at $P_{2}=P_{2max}$: the inversion saturates to 1, i.e. all atoms 
occupy the upper lasing level, the coherence between the lasing levels is 
destroyed, therefore the number of photons slows down and the lasing ceases altogether. Further 
increase of $P_{2}$ does not affect the picture since the medium became 
practically transparent.

\paragraph*{}
Finally, a comparison of threshold pump for the $\Lambda$ and V schemes indicates different dependence on the saturation parameter S. In the $\Lambda$ scheme the threshold pump $P_{1thr}$ increases by several orders of magnitude as S changes. In contrast, in the V scheme the threshold pump $P_{2thr}$ changes within one order of magnitude for the same range of changes of S. This difference can be seen by comparing Figs. 2 and 4a for the $\Lambda$ and V schemes respectively.

\paragraph*{}
In summary, we have shown that the dynamic behavior of lasers of $\Lambda$ 
and V - schemes of pumping is essentially different at both low and very high 
pump rates. In the case of $\Lambda$ configuration the photon number saturates 
with pumping. The threshold pump of $\Lambda$-scheme is determined by the 
spontaneous emission factor $\beta$. If $\beta$ is less than some minimal value the laser operates in a "quiet" regime when no lasing occurs. 

\paragraph*{}
In the V case the 
number of photons increases with pump up to some maximal value, then it begins to 
decrease. At the critical value of the pump rate the laser turns off because of 
destruction of the coherence between the lasing levels, which in turn determines 
the atomic lifetime. In this case the $\beta$ factor is not a fixed parameter, 
but depends upon the pump rate. 

\paragraph*{}
This paper was supported in part by the grant from the Israeli Ministry of 
Immigrant Absorption.

{\bf References}
\begin{enumerate}
\item\label{Feld} K. An et al., Phys. Rev. Lett. {\bf 73}, 3375 (1994).
\item\label{beta} There are many ways to calculate $\beta$-factor. The most recent 
approach can be found in M.P. van Exter, G. Nienhuis, and J.P. Woerdman, 
Phys. Rev. A {\bf 54}, 3553 (1996).
\item\label{RC} See, for example, P.R. Rice and H.J. Carmichael, Phys. Rev. A  
{\bf 50}, 4318 (1994) and references therein.
\item\label{Savage} Yi Mu and C.M. Savage, Phys. Rev A {\bf 46}, 5944 (1992).
\item\label{we1} G.A. Koganov and R. Shuker, E-print: quant-ph/9801045
\item\label{LWI} O. Kocharovskaya, P. Mandel, and Y.V. Radeonychev, Phys. Rev. A {\bf 45}, 1997 (1992); M.O. Scully, Phys. Rept. {\bf 219}, 191 (1992).
\item\label{Sigman} A.E. Sigman, Lasers (University Science Books, Mill Valley, 1986).
\item\label{gammas} L.A. Lugiato, Physica {\bf 81A}, 565 (1975); L.A. 
Pokrovsky, Teoertich. i Matematich. Fizika {\bf 37}, 102 (1978); T. 
Arimitsu and F. Shibata, J. Phys. Soc. Jpn. {\bf 52}, 772 (1983).
\item\label{Jin} R. Jin, D. Boggavarapu, M. Sargent III, P. Meystre, H.M. 
Gibbs, and G. Khitrova, Phys. Rev. A {\bf 49}, 4038(1994).
\item\label{Yamamoto} Y. Yamamoto and R.E. Slusher, Physics Today 46, 66 (1993).
\item\label{Hart,Kenedy} D.L. Hart and T.A.B. Kennedy, Phys. Rev. A {\bf 
44}, 4572 (1991).
\end{enumerate}

\includegraphics{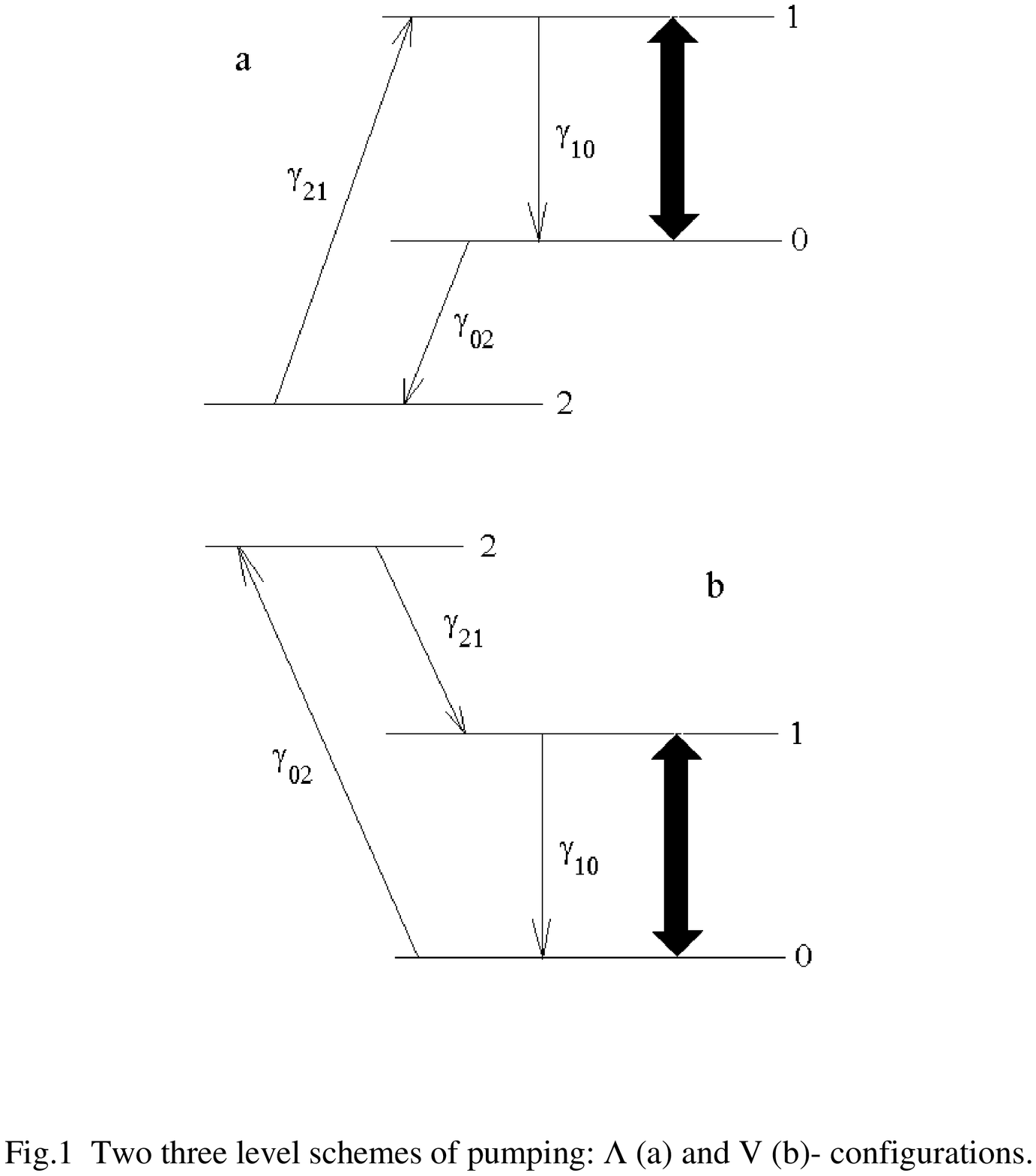}
\newpage
\includegraphics{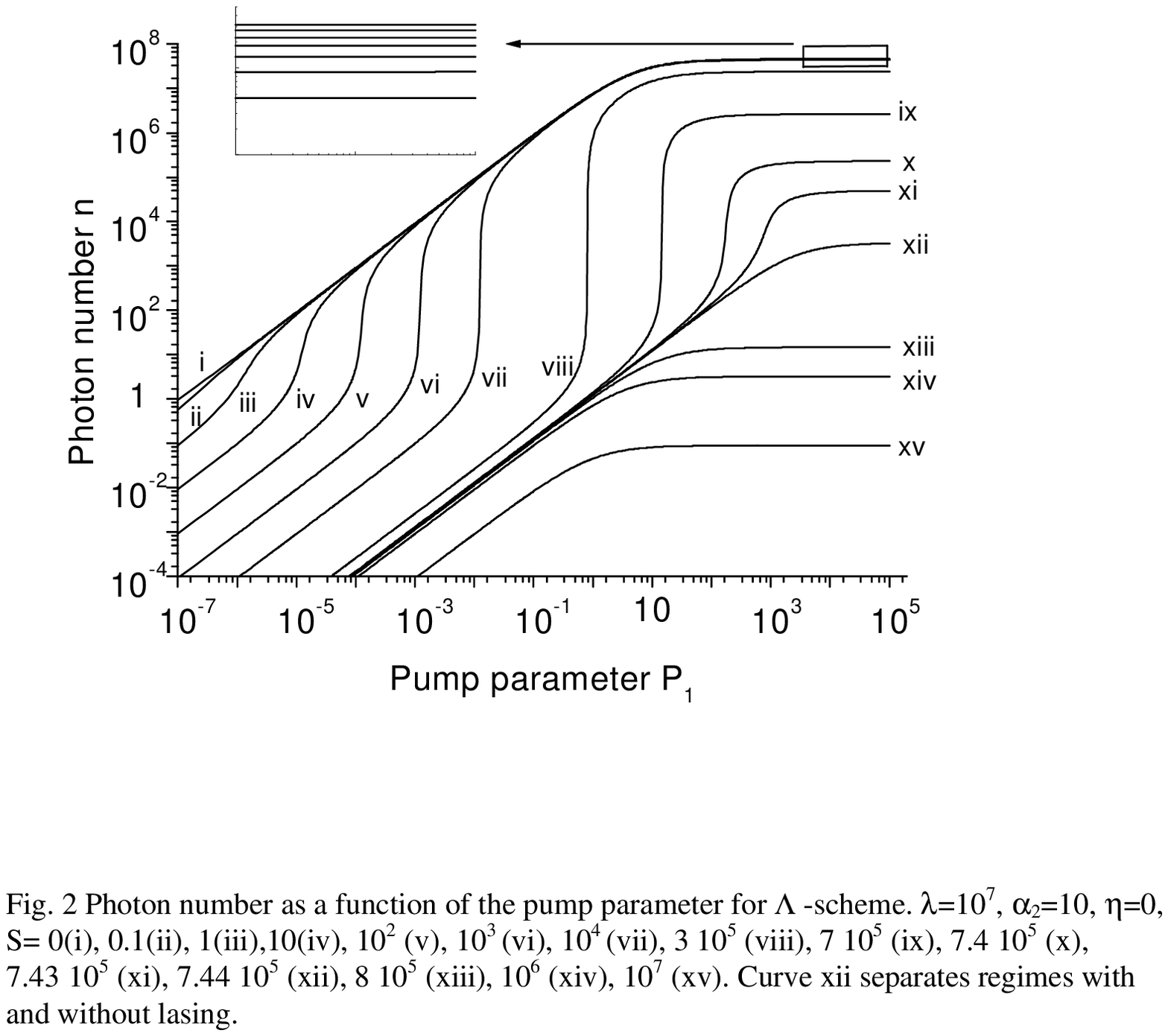}
\newpage
\includegraphics{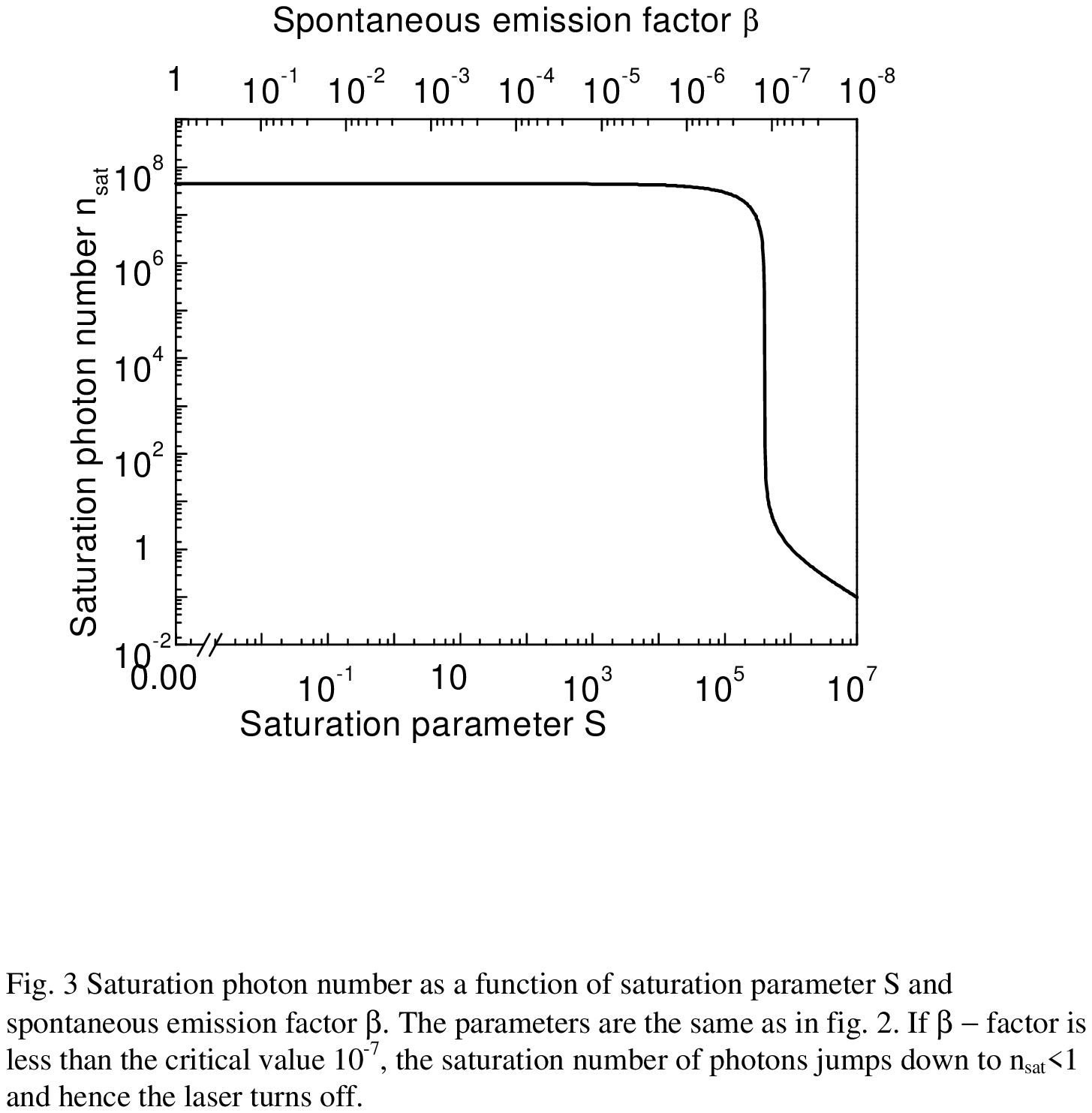}
\newpage
\includegraphics{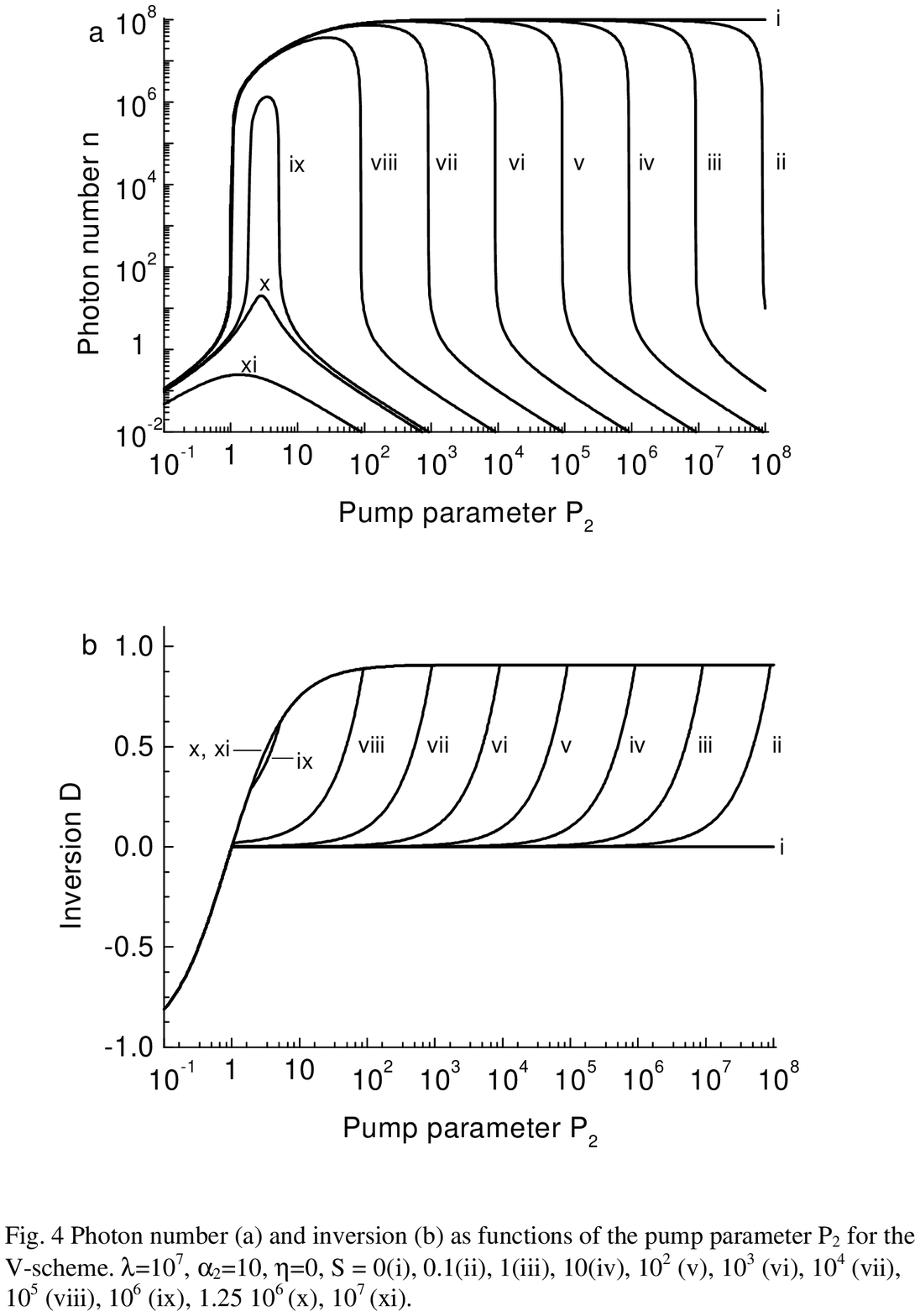}
\newpage
\includegraphics{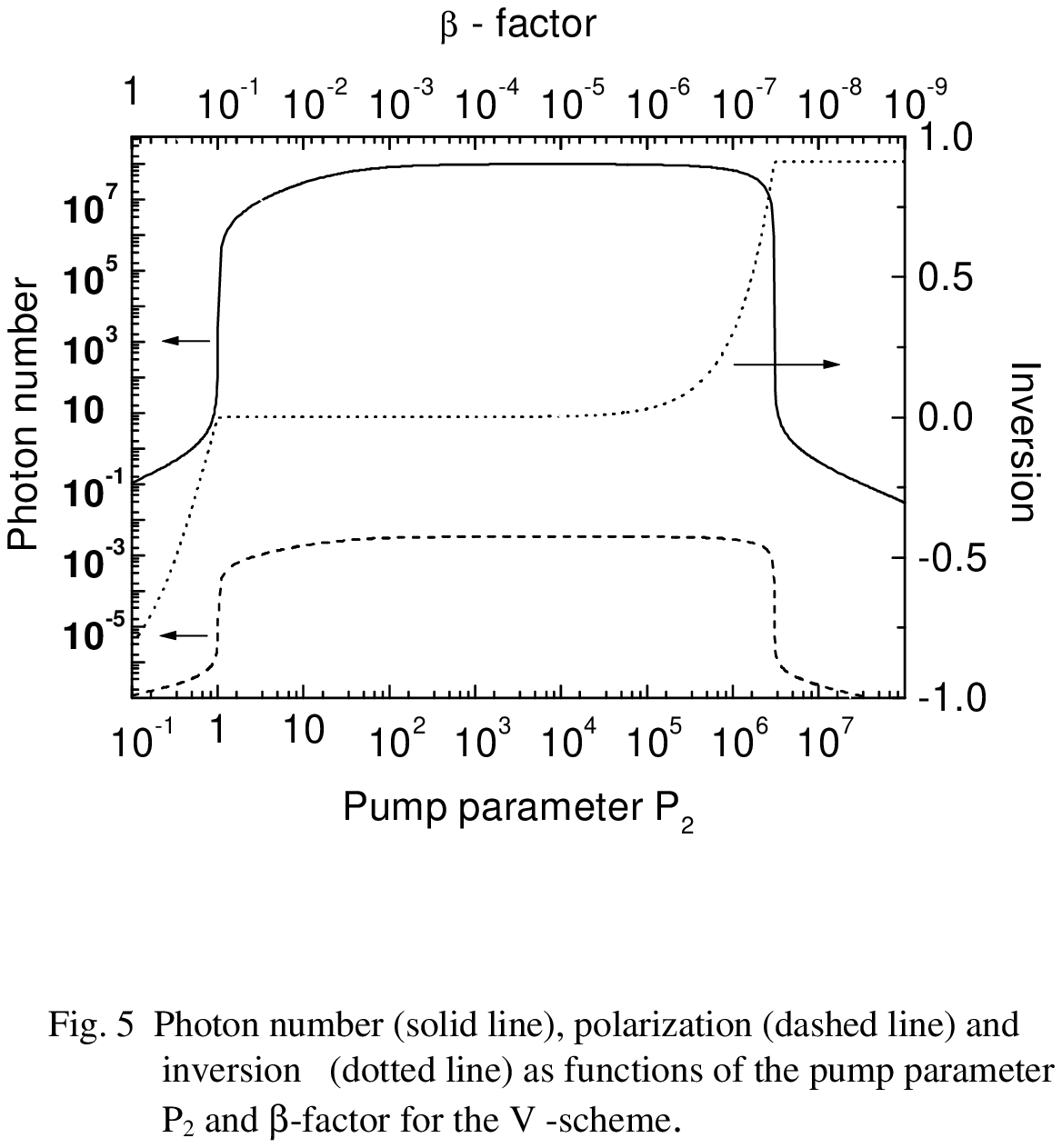}

\end{document}